**Hybrid photo-electrochemical and photo-voltaic cells (HPEV cells)**

*Gideon Segev[1,2], Jeffery Beeman[1,2], Ian D. Sharp[1,2,3]\**

[1]Chemical Sciences Division, Lawrence Berkeley National Lab, Berkeley, CA 94720, USA

[2]Joint Center for Artificial Photosynthesis, Lawrence Berkeley National Lab, Berkeley, CA 94720, USA

[3]Walter Schottky Institut and Physik Department, Technische Universität München, 85748 Garching, Germany

\*sharp@wsi.tum.deAbstract

The majority of photoelectrochemical (PEC) water splitting cells cannot drive the overall water splitting reactions without the assistance of an external power source. To provide added power, the cells are usually connected to photovoltaic (PV) devices in a tandem arrangement. This approach suffers from severe disadvantages since the PEC cell is connected in series to the PV cell and the overall current is typically limited by the saturation current of the PEC component. Thus, the operating point of the PV cell is often far from optimal and the overall system efficiency tends to be low. We propose a multi-terminal hybrid PV and PEC system (HPEV). As in tandem arrangements, the PEC cell is optically connected in series with the PV cell. However, a second back contact is used to extract the PV cell surplus current and allow parallel production of both electrical power and chemical fuel. Devices consisting of three-terminal silicon photovoltaic cells coupled to titanium dioxide water splitting layers are simulated and fabricated. The cells are shown to produce electricity with little reduction in the water splitting current, surpass the current mismatch limits, and increase the overall system efficiency.

## Introduction

The utilization of solar energy to drive electrochemical reactions is widely studied for energy storage, production of carbon-neutral fuels, environmental remediation, and other applications. However, in most solar-driven electrochemical devices, the potential required to drive the chemical reaction is much higher than the voltage that can be obtained from a high efficiency single junction solar cell. For this reason, there is a need for a tandem structure in which several materials are stacked in series each contributing some photovoltage. For example, in water spitting, although the Gibbs free energy is 1.23eV, the total photovoltage driving the reaction should be at least 1.5V because of overpotentials associated with the chemical reactions.[1,2] The optimal set of materials required to produce this voltage varies according to many parameters but in most cases, a combination of a silicon bottom junction with a wider band gap material yields the highest solar to hydrogen conversion efficiency[3] and is also the simplest from a technological standpoint. For this reason, significant efforts are undertaken aiming to realize water splitting devices where a wide band gap material, typically a metal oxide, serves as a top junction that is connected optically and electronically in series to bottom silicon junctions.[4–10] In such configurations the current through the system is determined by the least performing layer. This may be a result of low optical generation in a wide band gap material, lossy charge transport in a metal oxide, slow rate constants of the chemical reaction or a combination of some of them. As a result, nearly all reported devices operate at current densities that are far below the thermodynamic efficiency limit considering the solar energy input and the properties of the sub devices in the system.[4–8]

Here we propose a new class of devices, which can be classified as Hybrid Photo-Electrochemical and -Voltaic cells (HPEV cells). The HPEV cells overcome the problem of mismatched components performances by adding a third electrical terminal to the bottom junction. This third contact allows photogenerated charge carriers that are not consumed by the chemical reaction to be collected as electrical current, thereby producing electrical power at the same time that chemicals are produced. The functional performance of HPEV devices was validated through finite elements simulations. Next, proof of concept HPEV devices were fabricated and tested. The potential contribution of the HPEV for a wide range of PEC electrodes operated at different currents was studied with equivalent circuit modeling. We show that by collecting minority carriers that are not consumed by the chemical reaction as electrical current, electrical power can be harvested at the maximum power point with little effect on the chemical output. As a result, the overall system efficiency increases dramatically: a threefold increase in the overall performance can be achieved using state of the art photoanodes and back contact solar cells.

## Coupling losses in photoelectrochemical energy converters

The Gibbs free energy and overpotentials associated with chemical reactions define a minimum photovoltage that material stacks must generate to drive photoelectrochemical reactions. Figure 1a illustrates the performance of a water splitting device comprised of an ideal silicon bottom junction located behind a hypothetical PEC top junction. The band gap of the top junction is 2.1eV which is close to the band gap of widely studied photoanode materials such as hematite[11–14] or $Ta_3N_5$.[15–18] More details about the assumptions behind the performance of the two materials can be found in the supplementary information. The operating point of the integrated device is at the intersection of the two curves at current density and voltage $J_{op}$ and $V_{op}$ respectively. This point, which is marked by a circle in Figure 1a, corresponds to a current density of 12.45mA/cm² which yields a solar to hydrogen conversion efficiency of 15.3%. It can be easily noticed that this operating point is very far from the maximum power point of the silicon bottom junction (square in Figure 1a) demonstrating severe current mismatches (or coupling losses[19]). The current mismatch losses, $U_J$, can be defined as the power at the maximum power point reduced by the power at the operating point:

$$U_J = V_{mpp} \cdot J_{mpp} - V_{op} \cdot J_{op} \qquad (1)$$

where $J_{mpp}$ and $V_{mpp}$ are the current density and voltage at the maximum power point respectively. For the example illustrated in Figure 1a, the mismatch losses reach 13mW/cm² which translates to an absolute efficiency loss of 13%. This power is available for extraction but is not collected only because of poor integration between the two junctions. Nevertheless, the performance of the top junction in this example is far above all performance records demonstrated both in terms of extracted current and low onset potential.[12,14,15,20] This implies that any PEC solar water splitting system integrated with currently available materials will suffer from dramatically higher current mismatch losses. The stark contrast between the relatively high efficiencies that can be obtained by the silicon bottom junctions and the fractional amount of energy that is converted highlights the vast opportunities for efficiency increase in these systems.

Three terminal configurations have been suggested as means to remove coupling losses in a wide variety of multi-junction PV cells.[21–27] A detailed balance analysis of a transistor like, three terminal dual junction PV cell have shown that the efficiency limits for such devices is similar to that of standard dual junction solar cells.[28] However, in all of the suggested cells, the three terminal configuration is effectively realized by accessing a common contact between two sub- cells of opposite polarity. As a result, the current is split between the sub-cells and their voltage cannot be summed together. Hence, such configurations do not generate a sufficiently high voltage and a new cell design is required in order to implement the three terminal approach in PEC systems.

Figure 1b shows an illustration of a HPEV cell. In this example the bottom junction is made of a three-terminal silicon photovoltaic cell, on top of which the PEC electrode is deposited. The silicon cell bulk is n type doped and the photo-electrode operates as a photo-anode. The electrical coupling layer between the bottom junction and the PEC electrode, as well as underlayers and overlayers, are omitted from Figure 1b for simplicity. A photo-cathode based cell can be made with the opposite doping profiles. When photons are absorbed in the silicon bulk, electron hole pairs are generated (1). As in PV cells, the photo generated electrons flow to the n$^+$ back contact (2). Holes have two possible routes: they can flow towards the PEC contact from which they will be injected into the PEC layer and participate in the chemical reaction (3), otherwise, they can flow to the second, hole selective, back contact and contribute to the electrical power output (4). Hence, the back p$^+$ contact serves as an outlet for holes that are not injected into the top junction thus utilizing them for the generation of electrical power.

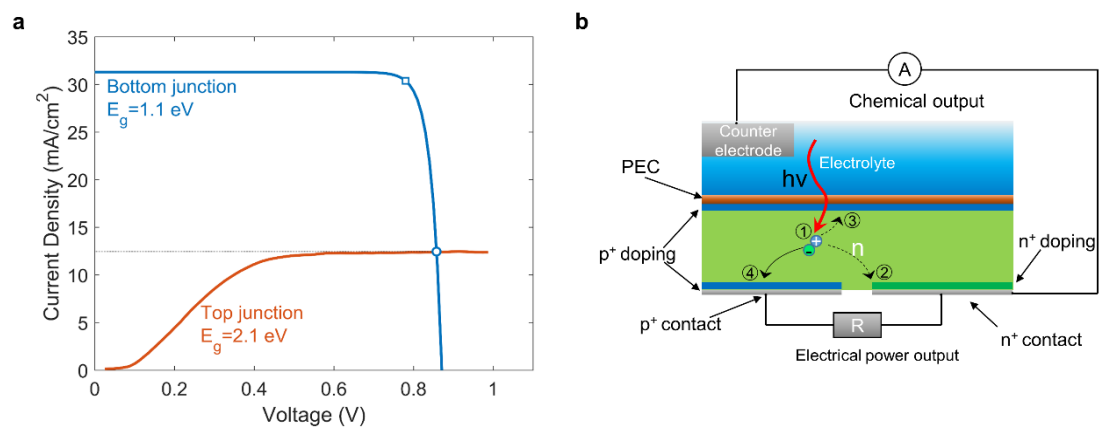

*Figure 1| Current mismatch losses and the HPEV cell. **(a)** Current voltage curves of an ideal 2.1eV band gap top junction and a Silicon bottom junction placed behind it. The integrated device operating point is at the intersection of the curves (circle). The maximum power point of the silicon bottom junction is marked with a square. **(b)** A schematic illustration of a HPEV cell.*

## Device simulation

The bulk of the three-terminal PV cell can be viewed as a reservoir of charge carriers. Within this reservoir, every charge carrier tends to flow towards the appropriate selective contact taking the path of least resistance depending on where it was photogenerated and the potential at the contact. For example, holes that are photogenerated near the top surface flow towards the top contact and are collected there if the operating point supports such currents. If the top contact is at a high voltage that can support only a fractional current, this will no longer be the path of least resistance and most of these holes will flow towards the back contact or recombine.

The performance of the HPEV cell is demonstrated with finite elements, solid state devices simulations (Sentaurus TCAD, Synopsys Inc.). We start by analyzing the performance of a three terminal silicon PV cell located behind a $TiO_2$ photoanode. The electrical coupling to a PEC photoanode will be done at a second stage as described below. More details about the simulation parameters and device geometry and can be found in the supplementary information.

At a first stage, the currents through the top and back $p^+$ contacts are calculated as a function of the voltage between the common ($n^+$) contact and the top contact, $V_1$, and the voltage between the two back contacts, $V_2$. Figure 2a shows the top and back $p^+$ contact currents as a function of the top contact voltage and several back $p^+$ contact voltages ($V_2$). Positive currents indicate that power is passed from the cell to the external circuit and negative currents indicate that power is passed from the external circuit into the cell. Since most of the photogenerated carriers are excited near the front surface, the path of least resistance for most holes is towards the top contact as long as the electrical operating point supports such current. Hence, when $V_2$=0 and the top contact voltage, $V_1$, is low, most of the current is extracted through the top contact. However, when increasing $V_1$, less current flows through it. As a result, collection through the back $p^+$ contact becomes the path of least resistance and the current through it increases. When $V_1$ passes the contact's open circuit voltage, the current through to top contact changes sign, i.e. the contact is injecting holes into the device. In a similar manner, when increasing $V_2$, less current is extracted through the bottom $p^+$ contact and the current top contact current increases. Hence, the currents through both top and bottom $p^+$ contact depend on both $V_1$ and $V_2$.

In full HPEV cell the three terminal PV cell coupled in series to a PEC top junction which determines the output of the top contact. In order to describe the performance of a spontaneous water splitting device, the current voltage curve of a $TiO_2$ photoanode as reported by Shaner et. al.[10] is used as the electrochemical load. Figure 2b shows the top contact current voltage curves for several back contacts voltages ($V_2$) and the electrochemical load curve defined by the $TiO_2$ photoanode. Since the operating point of the photoanode is the intersection of each current voltage curves with the electrochemical load, both the chemical and electrical operating points are controlled by the back contacts voltage. Nevertheless, changing the back contacts voltage from 0V to 0.6V shifts the operating point of the top junction by only 50mV hence the chemical output quite insensitive to the back contacts voltage.

Figure 2c shows the electrical and chemical outputs as a function of the back contacts voltage. The electrical output is the current drawn by the back contacts and the chemical output is the current density defining the reaction rate which is the point where the electrochemical load curve meets the top contact current voltage curve as shown in Figure 2b. The electric current extracted from the back contacts has little effect on the chemical output. For example, when the back contacts are short circuited, the chemical output is 0.405mA/cm$^2$ and when the back contacts are at the open circuit voltage, the chemical output is 0.49mA/cm$^2$. When the back contacts are at their maximum power point, the chemical output is 0.4175mA/cm$^2$. The inset shows the chemical and electrical power extracted from the device as a function of the back contacts operating point. The electrical power extracted is simply the product of the back P$^+$ contact current and voltage. Assuming a water splitting top junction, the chemical power output is the current through the top contact multiplied by 1.23V which is the Gibbs free energy for a water splitting reaction. At the maximum power point, the electric power output is 7.54mW/cm$^2$ and the chemical power output is 0.514mW/cm$^2$. When the back contacts are at open circuit voltage, i.e. no current is drawn back contacts, the chemical output is 0.585 mW/cm$^2$. Hence, a modest reduction of 71 μW/cm$^2$ in the chemical output allows extracting an electric power of 7.54mW/cm$^2$.

Similar to the use of point contacts in high efficiency silicon solar cells,[29,30] localized highly doped regions at the front surface and passivation of the free surfaces can reduce surface recombination losses dramatically. The dashed curves in Figure 2c describe the performance of an HPEV cell where the highly doped p$^+$ region covers a narrow 5μm wide region at the front surface instead of being a continuous layer along the entire surface. The removal of lossy, highly doped regions at the front surface increases the back contacts output dramatically yielding a short circuit current above 22.7mA/cm$^2$ which accounts for nearly 90% of the photogenerated charge carriers and a maximum extractable power of 11.2mW/cm$^2$. The reduced recombination also increases the voltage output of the front contacts which increases the chemical output to 0.535mW/cm$^2$ at the maximum power point. A significant increase in the device performance is also expected by improving the device optics for example by incorporating back reflectors and a textured front surface. Further optimization of the device geometry and doping profiles is left for future work.

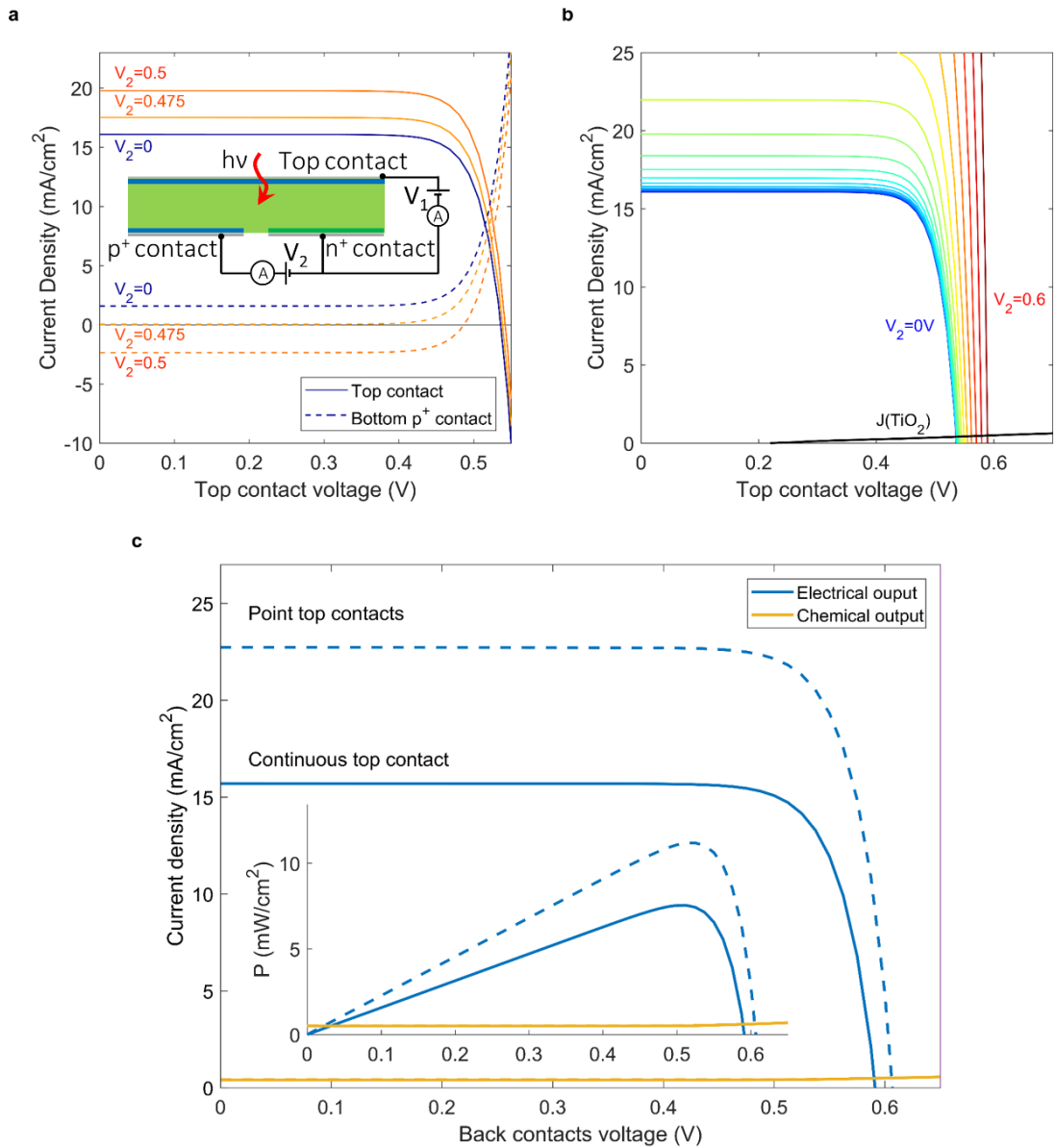

*Figure 2| Simulated HPEV performance. **(a)** The top and back p⁺ contacts currents as a function of the top contact voltage and several back p⁺ voltages. The inset shows a schematic of the simulated circuit. More details on the simulated geometry and material properties can be found in the supporting information. **(b)** The top contact current-voltage curves for several values of $V_2$. The black curve is the electrochemical load curve of the $TiO_2$ photoanode.[10] **(c)** Electrical and chemical outputs as a function of the back contacts voltage solid lines are for a continuous top contact and the dashed lines are for point top contacts. The inset shows the chemical and electrical power output (P) as a function of the back contacts voltage. Solid lines are for a continuous top contact and the dashed lines are for point top contacts.*

## Experimental validation

Demonstration HPEV cells were fabricated by depositing a $TiO_2$ water splitting layer on a three-terminal silicon solar cell as illustrated in Figure 1b. Details on the cell fabrication can be found in the methods section. At first stage, the electrical and chemical outputs were tested separately. The HPEV cell electric performance was tested by measuring the back contacts current voltage curves while the counter electrode was disconnected. The sample was illuminated with an AAA solar simulator and a UV LED was used to compensate for UV radiation content that is presents in the AM1.5G spectrum but missing from the solar simulator output. Details on the electrical characterization and the UV compensation can be found in the supporting information. Figure 3a shows the electrical output in the dark, and solar simulated light. The short circuit current density is 3.8 mA/cm$^2$ and the open circuit voltage is 0.44V. The maximum power point is at 0.32V with a power density of 1 mW/cm$^2$ and the fill factor is 0.62.

The PEC performance was tested in two and three electrodes configurations while the back p$^+$ contact was disconnected. Figure 3b and c show the PEC current voltage curves measured in three electrodes and two electrodes configurations respectively in the dark and under simulated 1 sun illumination. In both cases the working electrode was connected the n$^+$ back contact of the HPEV cell hence these curves include the photovoltage generated in the silicon. At 1.23V vs RHE the current density reaches 104µA/cm$^2$ under the simulated 1 sun illumination (Figure 3b) and 85µA/cm$^2$ under two electrodes, spontaneous water splitting conditions (Figure 3c).

Figure 3d shows the PEC and PV external quantum efficiency (EQE). Both PEC and PV were measured independently at short circuit. Details on the EQE measurement setup can be found in the methods section. The back contacts EQE was measured under several background light bias intensities as well as in the dark. The wavy features in the EQE spectra are a result interference patterns in the FTO and TiO$_2$ layers (The reflectance spectrum of the device is shown in Figure S11). The back contacts EQE increases with wavelength peaking at about 1000nm because lower energy photons are absorbed deeper in the silicon bulk and closer to the back contacts. The change in the back contacts EQE with the light bias indicates that the short circuit current is nonlinear with the light intensity. This effect was further studied by measuring and simulating the back contacts short circuit current under several light intensities. An elaborated discussion about the nonlinear dependence between the short circuit current and the light intensity can be found in the supplementary information.

The combined performance of the HPEV device was tested by measuring the electrical and spontaneous water splitting performances simultaneously as a function of the back contacts voltage. Details on how these measurements were conducted can be found in the methods section. Figure 3e shows the measured PV and PEC currents densities as a function of the back contacts voltage under several UV LED intensities thus probing the back contacts current as a function of the current extracted from the top contact. The black lines are the current responses to the simulated 1 sun spectrum. As predicted by the simulations, the PEC current is nearly independent of the electrical operating point allowing extraction of electrical power at the maximum power point without impeding the PEC output. As expected, the PEC current density increases with the LED power. Yet, the increase in PEC current has a minimal effect on the measured back contact current. Since the trajectories of charge carriers are defined by the path of least resistance, the PEC current is driven primarily by charge carriers that are generated near the top surface of the silicon. However, due to the short diffusion length (relative to the silicon wafer thickness), only carriers that are generated near the back surface are collected by the back contact. Hence, there is little competition over charge carriers between the front and back contacts and charge extraction off both front and back surfaces allows more current to be collected.

The fairly low power output of the fabricated devices is mostly because of the low short circuit current density due to short minority carriers lifetimes. Device simulations with a bulk lifetime of 15 µs and a surface recombination velocity of 1000cm/s yields a short circuit current density of 3.9 mA/cm$^2$ and show a good fit to the measured data (Figure S5). A similar bulk Shockley Read Hall lifetime was measured for wafers from the same batch with microwave photoconductance decay and by spatial collection efficiency extraction.[31] Hence, using high quality float zone silicon substrates, thickness optimization and passivation of the free surfaces at the back surface of the device provide a straight forward route for future efficiency enhancement. Other losses that can be reduced through device optimization include series resistance losses at the metal fingers and reduced open circuit voltage due to non-active areas at the device edges. Since these issues have already been solved by manufacturers of back contact solar cells, it can be expected that future performance of HPEV devices will be far higher than presented in this work.

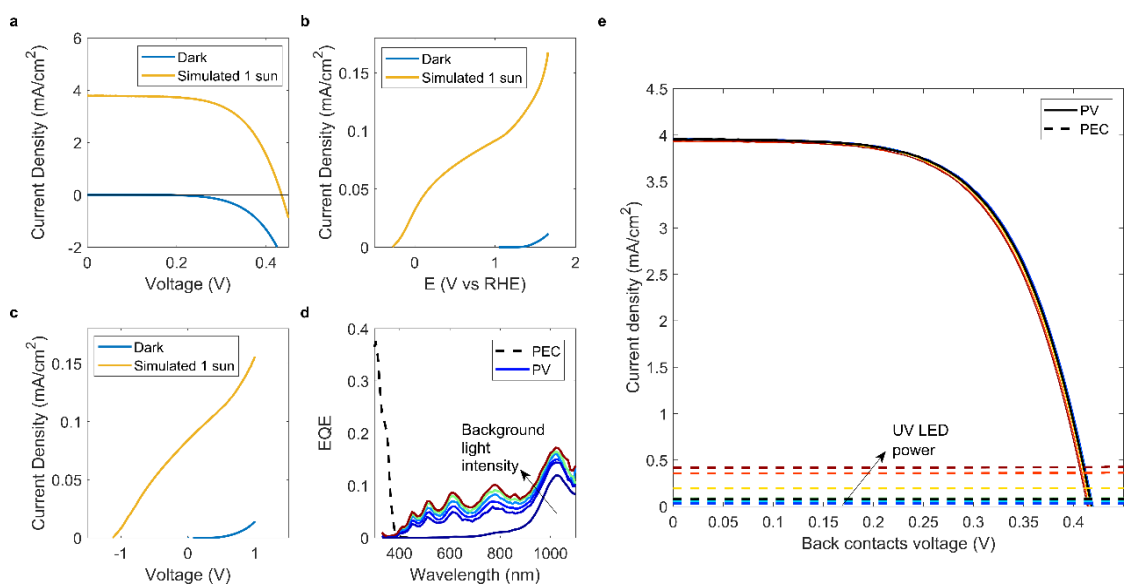

*Figure 3| HPEV cell characterization. Back contacts (**a**), three electrodes PEC (**b**) and two electrodes PEC (**c**) current voltage curves in the dark and under 1 sun simulated spectrum. The counter electrode was disconnected when measuring the back contacts' performance and the back p$^+$ contact was disconnected when measuring the photoelectrochemical performance. (**d**) The PEC and PV external quantum efficiency (EQE). Both PEC and PV are at short circuit. The back p$^+$ contact was disconnected when measuring the PEC and the counter electrode was disconnected when measuring the back (PV) contacts EQE. (**e**) PEC and PV currents as a function of the back contacts voltage for several UV LED power outputs. The black lines is the responses to the 1 sun simulated solar spectrum respectively.*

## Conceptual assessment

The potential contribution of the HPEV technology to the total energy produced by a solar fuels plant can be estimated through equivalent circuit modeling. Following the elaborated simulation and experimental results presented above, we assume that the PEC current is independent of the back contacts electrical operating point. In such case, an equivalent circuit of a simple solar cell can describe the HPEV back contacts output where the short circuit current is the overall current that is available for collection within the cell reduced by the PEC current. More details about the equivalent circuit model can be found in the supplementary information.

The back contacts current voltage curves and corresponding power output where calculated as a function of the top junction band gap and extracted PEC current density. The PV cell parameters were extracted out of the current voltage curve and EQE data reported by Mulligan et. al.[30] representing a standard performance for well-designed back contact solar cell. Figure 4 (a-c) show the HPEV calculated back contacts power output, chemical output and total efficiency as a function of the top junction current density and band gap respectively. The total efficiency is the sum of the chemical and electrical power outputs over the solar power input. The white region near the top right corner marks PEC current densities that cannot be reached considering the PEC layer band gap and the solar spectrum. In the white region next to the top left corner the PEC currents are limited by low photogeneration in the silicon. The star marks the BVO layer reported by Pihosh at. al.[20] with a band gap of 2.4eV and a current density of 5.57mA at the intersection between the PEC and solar cell current voltage curves. As can be expected, the back contacts power output is reduced as the PEC current increases and the PEC band gap is reduced. However, even for an ideal top junction, the back contacts produce significant amount of electrical power. For example, an ideal PEC with a band gap of 2.35eV generates a current density of 8.13mA/cm$^2$ (10% solar to Hydrogen efficiency). In this case the back contacts power output is 11.7mW/cm$^2$ and the total efficiency is nearly 22%. More generally, the electrical output makes at least half of the total power produced for any PEC material with a band gap of 2.3eV or above.

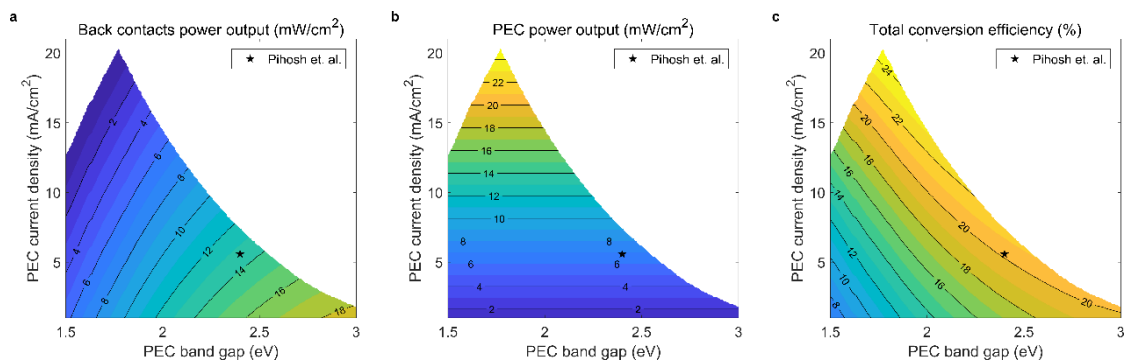

Figure 4|Equivalent circuit analysis of HPEV cells. The HPEV electrical power output (**a**), chemical output (**b**) and total efficiency (**c**) as a function of the PEC route current and photo-catalyst layer band gap. The star marks the BVO layer performance reported be Pihosh et. al.[20]

## Conclusion

A new class of devices, the hybrid photoelectrochemical and photovoltaic cells, were proposed. These devices are dual junction photoelectrochemical cells in which a second back contact is added to extract charge carriers that cannot be injected into the top junction due to current mismatches. The functional performance of the cells was studied with finite elements modeling and verified in prototypes fabricated from a silicon bottom junction and a TiO$_2$ top junction. It is shown that charge carriers that do not contribute to the chemical reaction can be harvested as electrical power at the maximum power point with a negligible effect on the chemical output. Equivalent circuit based modeling shows that HPEV cells made from off the shelf back contact solar cell can at least double the overall output of system for top junctions with band gaps above 2.3eV.

References

1.  Bak, T., Nowotny, J., Rekas, M. & Sorrell, C. C. Photo-electrochemical hydrogen generation from water using solar energy. Materials-related aspects. *Int. J. Hydrogen Energy* **27,** 991–1022 (2002).

2.  Walter, M. G. *et al.* Solar water splitting cells. *Chem. Rev.* **110,** 6446–73 (2010).

3.  Hu, S., Xiang, C., Haussener, S., Berger, A. D. & Lewis, N. S. An analysis of the optimal band gaps of light absorbers in integrated tandem photoelectrochemical water-splitting systems. *Energy Environ. Sci.* **6,** 2984–2993 (2013).

4.  Liu, C., Tang, J., Chen, H. M., Liu, B. & Yang, P. A Fully Integrated Nanosystem of Semiconductor Nanowires for Direct Solar Water Splitting. *Nano Lett.* **13,** 2989–2992 (2013).

5.  Shaner, M. R. *et al.* Photoelectrochemistry of core–shell tandem junction n–p$^+$-Si/n-WO$_3$ microwire array photoelectrodes. *Energy Environ. Sci.* **7,** 779–790 (2014).

6.  Chakthranont, P., Hellstern, T. R., McEnaney, J. M. & Jaramillo, T. F. Design and Fabrication of a Precious Metal-Free Tandem Core-Shell p$^+$n Si/W-Doped BiVO$_4$ Photoanode for Unassisted Water Splitting. *Adv. Energy Mater.* 1701515 (2017). doi:10.1002/aenm.201701515

7.  Jang, J.-W. *et al.* Enabling unassisted solar water splitting by iron oxide and silicon. *Nat. Commun.* **6,** 7447 (2015).

8.  Abdi, F. F. *et al.* Efficient solar water splitting by enhanced charge separation in a bismuth vanadate-silicon tandem photoelectrode. *Nat. Commun.* **4,** 2195 (2013).

9.  Sheridan, M. V. *et al.* All-in-One Derivatized Tandem p+n-Silicon-SnO2/TiO2 Water Splitting Photoelectrochemical Cell. *Nano Lett.* **17,** 2440–2446 (2017).

10. Shaner, M. R., McDowell, M. T., Pien, A., Atwater, H. A. & Lewis, N. S. Si/TiO$_2$ Tandem-Junction Microwire Arrays for Unassisted Solar-Driven Water Splitting. *J. Electrochem. Soc.* **163,** H261–H264 (2016).

11. Sivula, K., Le Formal, F. & Grätzel, M. Solar water splitting: progress using hematite (α-Fe$_2$O$_3$) photoelectrodes. *ChemSusChem* **4,** 432–49 (2011).

12. Warren, S. C. *et al.* Identifying champion nanostructures for solar water-splitting. *Nat. Mater.* **12,** 842–9 (2013).

13. Tilley, S. D., Cornuz, M., Sivula, K. & Grätzel, M. Light-induced water splitting with hematite: improved nanostructure and iridium oxide catalysis. *Angew. Chem. Int. Ed. Engl.* **49,** 6405–8 (2010).

14. Dotan, H. *et al.* Resonant light trapping in ultrathin films for water splitting. *Nat. Mater.* **12,** 158–64 (2013).

15. Liu, G. *et al.* Enabling an integrated tantalum nitride photoanode to approach the theoretical photocurrent limit for solar water splitting. *Energy Environ. Sci.* **9,** 1327–1334 (2016).

16. Takata, T. *et al.* Visible-light-driven photocatalytic behavior of tantalum-oxynitride and nitride. *Res. Chem. Intermed.* **33,** 13–25 (2007).

17. Li, Y. *et al.* Vertically Aligned $Ta_3N_5$ Nanorod Arrays for Solar-Driven Photoelectrochemical Water Splitting. *Adv. Mater.* **25,** 125–131 (2013).

18. Zhong, M. *et al.* Highly Active GaN-Stabilized $Ta_3N_5$ Thin-Film Photoanode for Solar Water Oxidation. *Angew. Chemie Int. Ed.* **56,** 4739–4743 (2017).

19. Rothschild, A. & Dotan, H. Beating the Efficiency of Photovoltaics-Powered Electrolysis with Tandem Cell Photoelectrolysis. *ACS Energy Lett.* 45–51 (2016). doi:10.1021/acsenergylett.6b00610

20. Pihosh, Y. *et al.* Photocatalytic generation of hydrogen by core-shell WO3/BiVO4 nanorods with ultimate water splitting efficiency. *Sci. Rep.* **5,** 11141 (2015).

21. Tai, C.-H., Lin, C.-H., Wang, C.-M. & Lin, C.-C. Three-Terminal Amorphous Silicon Solar Cells. *Int. J. Photoenergy* **2011,** 1–5 (2011).

22. Sista, S., Hong, Z., Park, M. H., Xu, Z. & Yang, Y. High-efficiency polymer tandem solar cells with three-terminal structure. *Adv. Mater.* **22,** 77–80 (2010).

23. D??rr, M., Bamedi, A., Yasuda, A. & Nelles, G. Tandem dye-sensitized solar cell for improved power conversion efficiencies. *Appl. Phys. Lett.* **84,** 3397–3399 (2004).

24. Nagashima, T., Okumura, K., Murala, K. & Kimura, Y. Three-terminal tandem solar cells with a back-contact type bottom cell. *Conf. Rec. IEEE Photovolt. Spec. Conf.* **2000–January,** 1193–1196 (2000).

25. Guo, F. *et al.* A generic concept to overcome bandgap limitations for designing highly efficient multi-junction photovoltaic cells. *Nat. Commun.* **6,** 7730 (2015).

26. Bahro, D. *et al.* Understanding the External Quantum Efficiency of Organic Homo-Tandem Solar Cells Utilizing a Three-Terminal Device Architecture. *Adv. Energy Mater.* **5,** 1–8 (2015).

27. Adhyaksa, G. W. P., Johlin, E. & Garnett, E. C. Nanoscale Back Contact Perovskite Solar Cell Design for Improved Tandem Efficiency. *Nano Lett.* **17,** 5206–5212 (2017).

28. Martí, A. & Luque, A. Three-terminal heterojunction bipolar transistor solar cell for high-efficiency photovoltaic conversion. *Nat. Commun.* **6,** (2015).

29. Zhao, J., Wang, A. & Green, M. a. 24·5% Efficiency silicon PERT cells on MCZ substrates and 24·7% efficiency PERL cells on FZ substrates. *Prog. Photovoltaics Res. Appl.* **7,** 471–474 (1999).

30. Mulligan, W. W. P. *et al.* Manufacture of solar cells with 21% efficiency. *Proc. 19th EPVSEC* 3–6 (2004).

31. Segev, G. *et al.* The spatial collection efficiency of photogenerated charge carriers in photovoltaic and photoelectrochemical devices. *Joule* **Accepted,** (2017).

32. Shockley, W. & Queisser, H. J. Detailed Balance Limit of Efficiency of p-n Junction Solar Cells. *J. Appl. Phys.* **32,** 510–519 (1961).

33. Chung, D., Davidson, C., Fu, R., Ardani, K. & Margolis, R. *U.S. Photovoltaic Prices and*

*Cost Breakdowns. Q1 2015 Benchmarks for Residential, Commercial, and Utility-Scale Systems*. (2015). doi:10.2172/1225303

34. Yoshikawa, K. *et al.* Silicon heterojunction solar cell with interdigitated back contacts for a photoconversion efficiency over 26%. *Nat. Energy* **2,** 17032 (2017).

35. Warren, S. C. *et al.* Identifying champion nanostructures for solar water-splitting. *Nat. Mater.* **12,** 842–9 (2013).

# Hybrid photo-electrochemical and photo-voltaic cells (HPEV cells)- supplementary information

*Gideon Segev[1,2], Jeffery Beeman[1,2], Ian D. Sharp[1,2,3]**

[1]Chemical Sciences Division, Lawrence Berkeley National Lab, Berkeley, CA 94720, USA

[2]Joint Center for Artificial Photosynthesis, Lawrence Berkeley National Lab, Berkeley, CA 94720, USA

[3]Walter Schottky Institut and Physik Department, Technische Universität München, 85748 Garching, Germany

[*]sharp@wsi.tum.deAbstract

This file includes:

    Methods

    Supplementary Information

    Supplementary references

Methods

## 1. Device fabrication

The devices used in this study were fabricated using 2-side polished, 5-10Ωcm, Phosphorus-doped, 0.25mm thick Silicon wafers that were <1,0,0> Czochralski grown. The light-incident side of these wafers was ion implanted with Boron in a 2-step fashion: 3 x $10^{14}$ cm$^{-2}$ at 33 keV, followed by 5 x $10^{14}$ cm$^{-2}$ at 50 keV. This treatment produces a reasonably uniform, metallically-doped (5 x $10^{18}$ – 5 x $10^{19}$ cm$^{-3}$) contact region that extends 250 nm from the surface into the bulk. The backside of the samples received both n$^+$ and p$^+$ implanted contacts in an interleaved comb pattern structures where each finger is 60μm wide with a 100μm spacing between fingers, as shown in Figure S5. The Phosphorus, n$^+$ type, ion implants dosage was 2 x $10^{14}$ cm$^{-2}$ at 33 keV, followed by 5 x $10^{14}$ cm$^{-2}$ at 75keV. The Boron p$^+$ type implants dosage was 3 x $10^{14}$ cm$^{-2}$ at 33 keV, followed by 5 x $10^{14}$ cm$^{-2}$ at 50 keV. After photoresist stripping and oxide removal using an HF vapor etch, the implanted atoms were activated using a 900° C, 10 second rapid thermal anneal. After the dopants implantation and activation, a fluorine doped tin oxide (FTO) layer was deposited on the top surface using Ultrasonic Spray Pyrolysis (USP) followed by an electron beam evaporation of TiO$_2$ and post annealing. A final photolithographic step and e-beam evaporator was then used to overlay the backside, comb-patterned ion implants with metal contacts. We used 20nm of Titanium followed by 300 nm of gold for this purpose. The width of the metal fingers is XX, slightly thinner than the highly doped regions, to avoid shunting of the highly doped regions and the device bulk.

Following metallization, the wafers were diced into 12 mm×11 mm chips each with a single HPEV device as illustrated in Figure S5.

## 2. FTO deposition

Fluorine doped tin oxide was deposited with Ultrasonic Spray Pyrolysis (Sono-Tek ExactaCoat) on freshly etched surfaces (1min 5% HF). The heating plate temperature was set to 500 °C and the flow was set to 1mL/min. The spraying speed was set to 100 mm/s with 20 repetitions resulting in a FTO layer thickness of about 200nm (see cross section in Figure S10). The precursor used for the depositions was made by mixing of 90.2 mL Ethanol and 7.23 mL Butyltin Tricholride (Aldrich 95%)with a solution containing 0.122 g Amonium Fluoride (Aldrich 99.99%) in 2.46 mL water.

## 3. TiO$_2$ deposition

The PEC layer was deposited using electron beam evaporation (Angstrom Engineering, NEXDEP) of TiO$_2$ (Kurt J. Lesker, Titanium Dioxide pieces 99.9% pure). The deposition was conducted at a substrate temperature of 350 °C at a vacuum of about $10^{-6}$ Torr. The acceleration voltage was set to 7 kV and the deposition rate to 0.5Å/s. Post deposition air annealing was done at 500°C for 3.5 hours. A plan and cross sectional views of the TiO$_2$ layers can be seen in Figure S10 (a-b) respectively.

## 4. EQE measurements

External quantum efficiency (EQE) measurements were carried out using a Newport 300 W ozone-free Xe lamp, whose optical output was passed through an Oriel Cornerstone 130 1/8 m monochromator. The sample current was measured with a Gamry Reference 600 potentiostat. The monochromatic light was stepped in 5 nm intervals and chopped at a period between 0.5 and 5 seconds depending on the settling time of the current signal. A Mightex GCS-6500-15-A0510 light emitting diode and a Mightex LGC-019-022-05-V collimator were used to produce the background light bias. Back contacts EQE measurements were conducted with several background light intensities with LED current of 25mA, 50mA, 100mA, 200mA, 400mA which corresponds to background current densities of 81, 212, 518, 1150 and 2350 µA/cm$^2$ respectively and without the light bias. The PEC counter electrode was disconnected during these meausrements. The PEC EQE was conducted in two electrodes and in short circuit while the back p$^+$ was disconnected. The background light bias with an LED current of 500mA. The photocurrent was calculated by subtracting the current generated under background light illumination from the current generated in the presence of both monochromatic and background light illumination. The incident optical output at each wavelength was measured with a Thorlabs SM05PD2A photodiode. The photodiode was calibrated using a Newport 818-UV/DB calibrated detector.

## 5. Photoelectrochemical characterization

Photoelectrochemical characterization was conducted using an AAA solar simulator (Oriel Sol3A 94023A) and  A 340nm collimated LED (Thorlabs M340L4 with Thorlabs COP1A collimating optics). The chemical output was measured in a 1M NaOH electrolyte with a Bio-Logic VSP potentiostat in a two electrodes configuration using a platinum wire as counter electrode and in a three electrodes configuration with a platinum wire counter electrode and a leak-free Ag/AgCl reference electrode. The electrical output was measured using the second channel of the potentiostat. Simultaneous water splitting and electric power generation measurements where conducted by connecting each output to a different potentiostat channel and synchronizing the two channels. A linear sweep voltammetry was conducted in order to measure the electrical output while a two electrodes chronoamperometry measurement at a potential of 0V versus the reference was used to monitor the chemical output.

# Supplementary information

## 1. Simulation parameters and geometry

The current voltage curves shown in Figure 1 presents a combination an ideal photovoltaic bottom junction with a band gap of 1.1eV and an ideal top PEC junction with a band gap of 2.1eV. The bottom junction current voltage curve is calculated according to the detailed balance limit for this combination of band gaps.[1] The current voltage curve for the top PEC junction is drawn schematically and it illustrates the uppermost PEC performance limits by assuming that all available above band gap photons are harvested as current at the operating point.

Detailed simulations of the HPEV functional performance were carried out through finite elements simulations. The simulation tool (Sentaurus TCAD, Synopsys Inc.) solves the coupled Poisson and continuity equations for electrons and holes under boundary conditions that determine the electrical operating point and an optical generation profile produced by the standard AM 1.5 solar spectrum.

The optical generation is calculated using the transfer matrix method (TMM) where the cell is illuminated through the top contact surface with a standard solar spectrum AM 1.5G where only wavelengths that are not absorbed in the top junction are considered. The simulated device is a three terminal Si solar cell similar to the illustration shown in Figure 1b. The bulk is $10^{15}$cm$^{-3}$ doped n type and the common contact is through a highly doped n$^+$ region located on the back surface. A second contact is made through a p$^+$ region also at the back surface of the cell and the third terminal is through another p$^+$ layer covering the entire top surface (facing the sun). All highly doped regions are doped at a concentration of $10^{19}$ cm$^{-3}$. The bulk maximum Shockley Read Hall lifetime is 1ms and the surface recombination velocity of free surfaces is 10cm/s unless stated otherwise. Figure S1 shows the simulated device geometry and Table S1 lists all the simulation parameters.

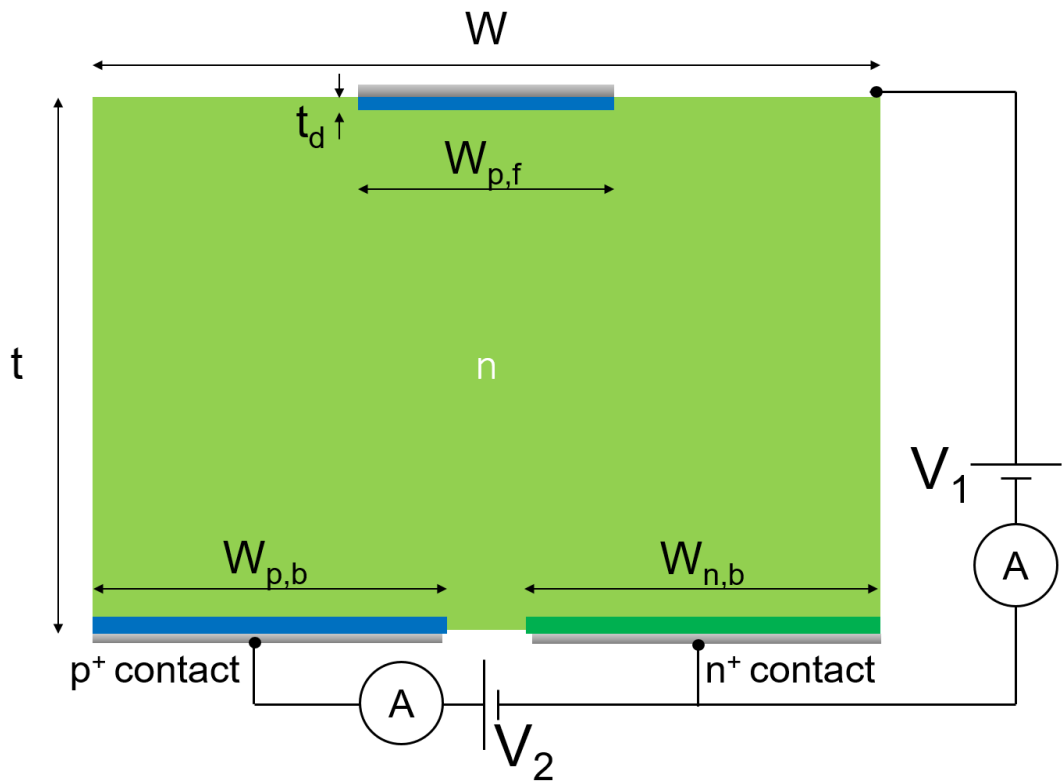

*Figure S1: The simulated device geometry and electrical configuration.*

| Symbol | Description | value |
| --- | --- | --- |
| W | Cell width | 100 μm |
| t | Cell thickness | 250 μm |
| $W_{p,b}$ | Back p$^+$ region width | 30 μm |
| $W_{n,b}$ | Back n$^+$ region width | 30 μm |
| $W_{p,f}$ | front p$^+$ region width (only in point contacts simulations) | 5 μm |
| $t_d$ | Highly doped regions thickness | 3 μm |
| $N_b$ | Bulk doping | $10^{15}$ cm$^{-3}$ |
| N$^+$ | n$^+$ region doping | $10^{19}$ cm$^{-3}$ |
| P$^+$ | p$^+$ regions doping | $10^{19}$ cm$^{-3}$ |
| SRV | Surface recombination velocity | 10 cm/s |
| $\tau_{n,p}$ | Electrons and holes maximum bulk lifetime | 1 ms |

*Table S1: parameters used in the finite elements simulations.*

## 2. Simulated current streamlines

Figure S2 a-c shows streamlines of the current flow through the 3 terminal PV cell under different biases: all contacts are short circuited (a), the back p[+] contact is at $V_2$=0.55V and the top contact is at $V_1$=0 (b), and the top contact is at $V_1$=0.55V and the back p[+] contact is at $V_2$=0 (c). The device geometry and material properties are as in Figure 2. The color coding shows the potential distribution within the cell. Similar Figure 1b, the common contact is located near the bottom right corner, and back p[+] contact is near the bottom left corner of the device. An illustration of the simulated device and circuit is shown in the inset in Figure 2a. When all contacts are short circuited (Figure S2a), current from the common n[+] contact is split between the two other contacts. More current is driven through the top contact because most of the minority charge carriers are generated next to it. Hence, collection through the top surface is the path of least resistance for most photogenerated holes. This tendency has an important effect on the full HPEV cell as described in the device simulation section. When the top contact is at $V_1$=0V and the back p[+] contact is at $V_2$=0.55V, which is slightly above its open circuit voltage (Figure S2b), current from both back contacts flows towards the top contact. When the top contact is at $V_1$=0.55V and $V_2$=0V, current is injected from the top contact into the cell and the sum of the top contact and the common contact currents is extracted through the back p[+] contact as shown in Figure S2c.

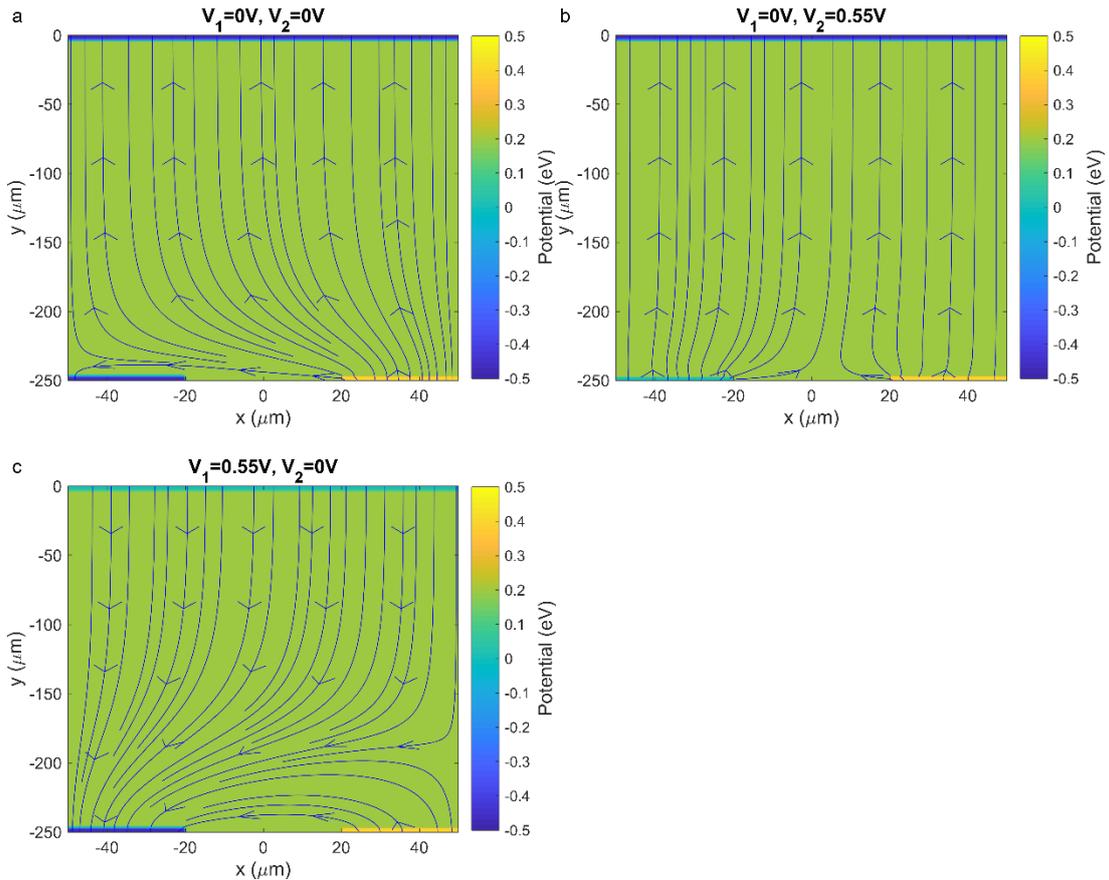

*Figure S2: Streamlines showing current flow through a 3 terminal solar cell under different biases: all contacts are short circuited (a), the top contact is at $V_1$=0V and the back p[+] contact is at $V_2$=0.55V (b), and the top contact is at $V_1$=0.55V and the back p[+] contact is at $V_2$=0V (c). The color coding shows the potential distribution within the cell. The common contact is located at near the bottom right corner, and back p[+] contact is near the bottom left corner of the device.*

## 3. Device optimization

Although the overall efficiency of the simulated and fabricated HPEV cells is dramatically higher than the efficiency of standard PEC cell made with the same components, the power collected by the back contacts still falls short of what is obtainable with state of the art back contact solar cells.[2,3] The lower efficiency obtained by the simulated devices is a result of non-optimal geometry and optics. As discussed in the main text, in the point contact geometry, nearly 90% of the absorbed photons are collected in short circuit conditions. Since no anti reflection coating is assumed, about 30% of the incident light is reflected off the front surface. Thus, significant improvement is expected by front surface texturing or other optical management schemes. Next steps towards higher efficiency should address the open circuit voltage. Optimization of the back contacts magnitudes and pitch may reduce recombination in these regions thus increasing the open circuit voltage.

In standard photovoltaic and PEC cells the goal of the device optimization process is usually to maximize the overall power output of the device. In HPEV devices such optimization may be conceptually different since an increase in one output may result in a decrease of the other. For example, nanostructuring is widely used to enhance the absorptance and reduce the distance carriers must travel in PEC cells.[4–6] However, since such structures increase light scattering, they might reduce light absorption in the silicon bottom junction and its electrical power output. Similar tradeoffs can be encountered when optimizing the doping profiles of top point contacts, the TCO layer thickness and other elements in the device geometry. Hence, in this class of devices, similar to other systems in which there is more than one possible output, optimization must be done in light of a clear objective function that defines the desired ratio of products for the specific application. For example, in standalone systems, where the electric power output is used only to power peripheral components such as compressors and sensors, the chemical output can be maximized at the expense of electricity production. On the other hand, the optimal ratio between the chemical and electric output may change during the course of a day when electricity is sold to the grid leading to a different optimal cell design.

## 4. Effects of bulk lifetime and surface recombination velocity

In all the simulations presented in the main text, the minority carriers lifetime and surface recombination velocities describe a device with state of the art material properties and superb surface passivation. Lower minority carriers lifetimes and higher surface recombination will impede the device performance requiring different optimization strategies. Figure S3 a shows the HPEV back contacts current voltage curves for several bulk lifetimes. The surface recombination velocity is 1000 cm/s. Figure S3 b shows the HPEV back contacts current voltage curves for several surface recombination velocities and a bulk lifetime of 1ms. All other material propterties as well as the PEC layer properties (not shown) are as in Figure 2. As in back contacts photovoltaic cells, the material properties have a determinental effect of the device performance. Since minority carriers must traverse most of the device thickness before being collected, the short circuit current of HPEV cells is very sensitive to the minority carriers lifetime. Similarly, high surface recombination will drive minority carriers towards the non active surface thus cancelling the net driving froce towards the back contacts.

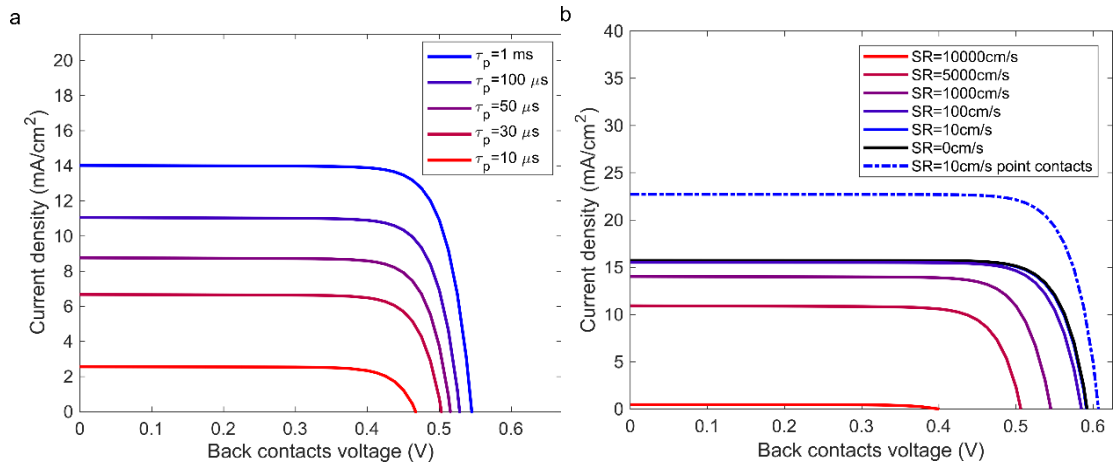

*Figure S3: Back contacts current voltage curves for several bulk lifetimes and a surface recombination velocity of 1000cm/s (a). Back contacts current voltage curves for several surface recombination velocities and a bulk lifetime of 1ms (b). All other material propterties as well as the PEC layer properties are as in Figure 2.*

## 5. Fabricated devices geometry

The interleaved n⁺ and p⁺ doped regions at the back surface of the cells were implanted through two designated photolithographic masks producing an interdigitated structure as illustrated in Figure S4. HPEV devices were fabricated on 4 inch wafers which were diced to produce 12 mm by 11 mm chips each containing a single HPEV device. The active area for the electrical output cell is defined as the area of the interlaced fingers at the back contacts of the cell. Figure S5 shows a schematic of the geometry of the back of the fabricated cells. It should be noted that the none active regions in the chip can effectively function as dark diodes that are connected in parallel to the active area thus reducing the voltage that can be extracted from the cell. This effect can be reduced by having larger cells with smaller edges and contact pads.

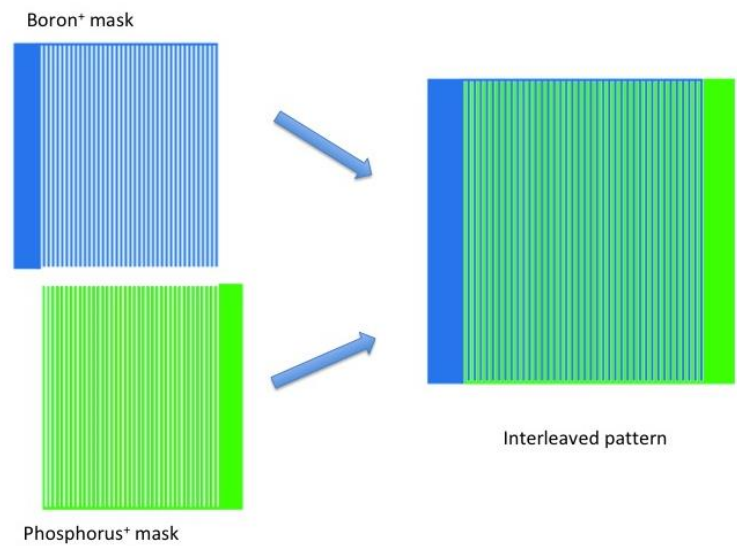

*Figure S4: Photolithographic masks were used to produce interleaved p⁺ and n⁺ contacts.*

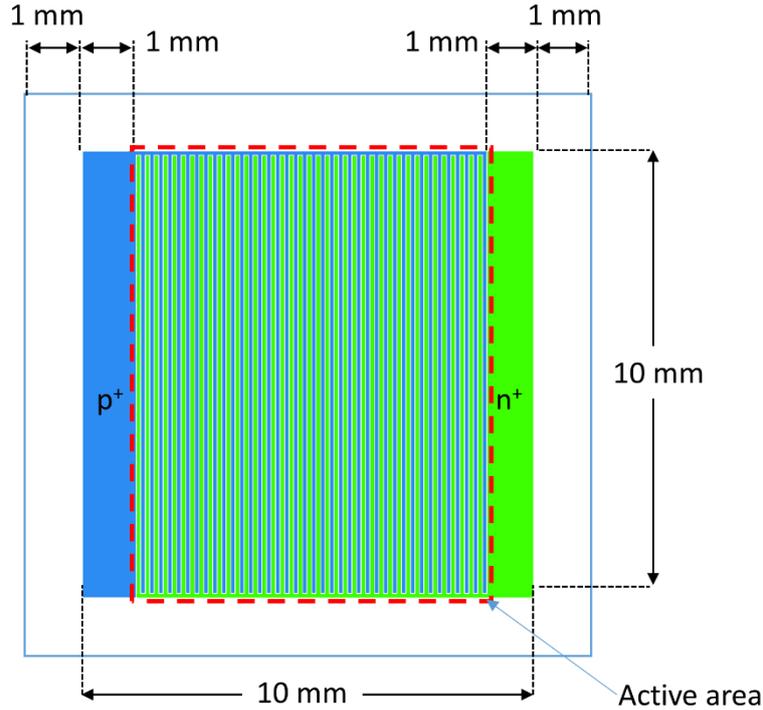

*Figure S5: The fabricated devices geometry. The white square is the total chip magnitude, the blue and green regions are the interlaced p⁺ and n⁺ type doped regions. The active area magnitude is 8mm by 10mm and the chip size is 11 mm by 12 mm.*

### 6. High light intensity measurement

Electrical characterization under high light intensity was carried out by positioning a Fresnel lens (Thorlabs FRP251) between the solar simulator and the device. The output intensity was controlled with the solar simulator internal shutter and the output was measured with a reference photodiode (Newport 91150V Reference cell and meter). Since the Fresnel lens absorbs a considerable portion of the UV content, the chemical output is significantly lower in these measurements.

Figure S1 shows the measured short circuit current density per sun as a function of the flux concentration. Also shown in Figure S1 are the simulated back contacts short circuit current densities per sun with a top contact current of 0.1mA/cm$^2$ and when the top contact is open circuited. The simulated device has a bulk lifetime of 15µs and a surface recombination velocity of 1000cm/s. The good agreement between the simulation results and measurements indicate that the none-linear relation between the current density and light intensity, as seen in figure 3d, is a basic aspect of these devices. Increasing the light intensity reduces the band bending near the front surface and with it driving forces for holes towards the front surface. As a result, more holes can reach the back contacts and contribute to the collected current. Further optimization is required in order to maximize the current collected by the back contacts which becomes linear with light intensity when increasing the bulk lifetime.

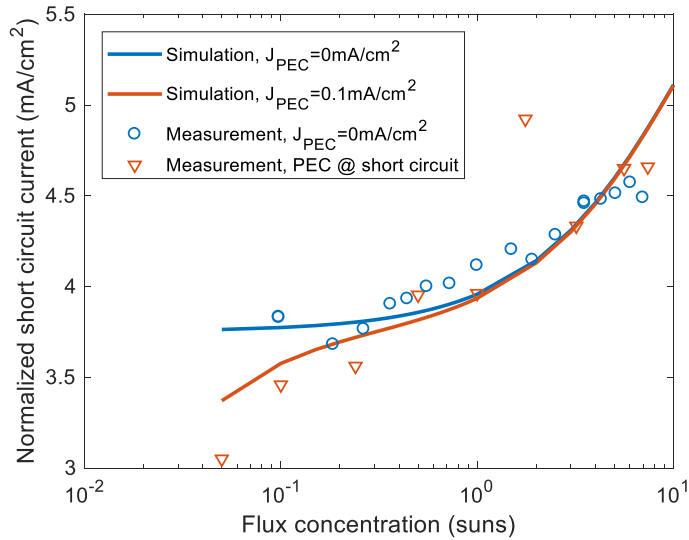

*Figure S1: The simulated and measured short circuit current density per sun as a function of the light intensity. Measured and simulated with the counter electrode disconnected ($J_{PEC}$=0mA/cm$^2$) and under combined PV and PEC operation.*

## 7. UV compensation

Since TiO$_2$ has a very wide band gap, it utilizes only photons within the UV portion of the solar spectrum. However, AAA solar simulators that are widely used to characterize PV and PEC cells are usually not optimized to match the standard AM1.5G spectrum in this range. Figure S6 shows the AAA solar simulator (Oriel Sol3A 94023A) and the AM 1.5G reference spectra within at wavelengths below 380nm. The inset shows the same spectra within the UV, visible and near IR range. The spectrum was measured with an ocean optics spectrometer (Ocean Optics HR2000+ES, factory calibrated for the 200-1050nm range). The overall intensity was measured to be 1 sun using a reference photodiode (Newport 91150V Reference cell and meter). As can be seen in Figure S6, although the reference spectrum has an insignificant number of photons at wavelengths below 360nm, the output of the solar simulator is very low in this range. For this reason, a 340nm UV LED (Thorlabs M340L4 with Thorlabs COP1A collimating optics) was used to compensate for the UV photons missing in the solar simulator output. The intensity of the LED was tuned such that the overall number of photons produced by the solar simulator and the LED will be the same as the AM 1.5G spectrum below 380 nm. The red curve in Figure S6 shows the UV compensated spectrum . This spectrum is referred to as the simulated 1 sun spectrum throughout this work. It should be noted that since all the photons below a wavelength of 380nm are absorbed in the TiO$_2$ layer, the added UV content has no effect on the HPEV back contact performance. Figure S7 shows the HPEV cell back contacts current voltage curves in the dark and when illuminated with the AAA solar simulator and simulated 1 sun spectra.

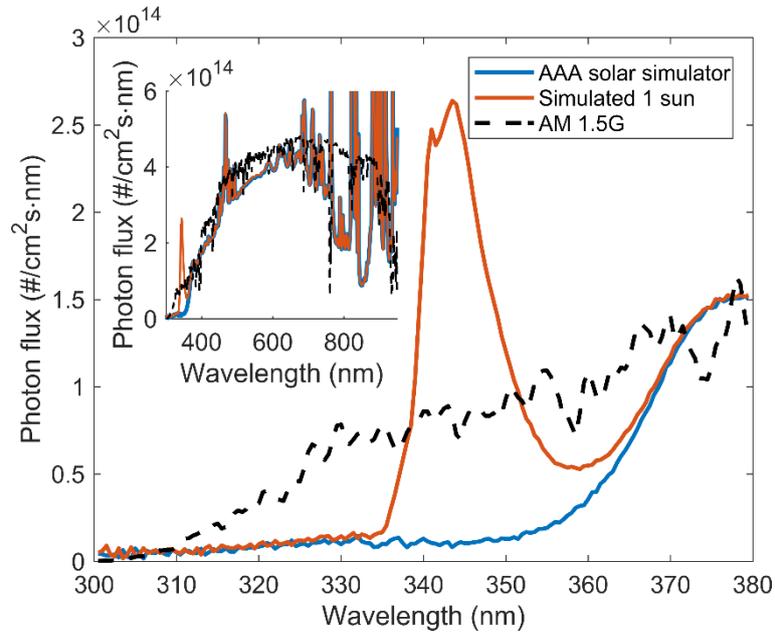

*Figure S6: The AAA solar simulator, simulated 1 sun and the AM 1.5G spectra within the UV range. The inset shows the same spectra in the UV, visible and near IR range.*

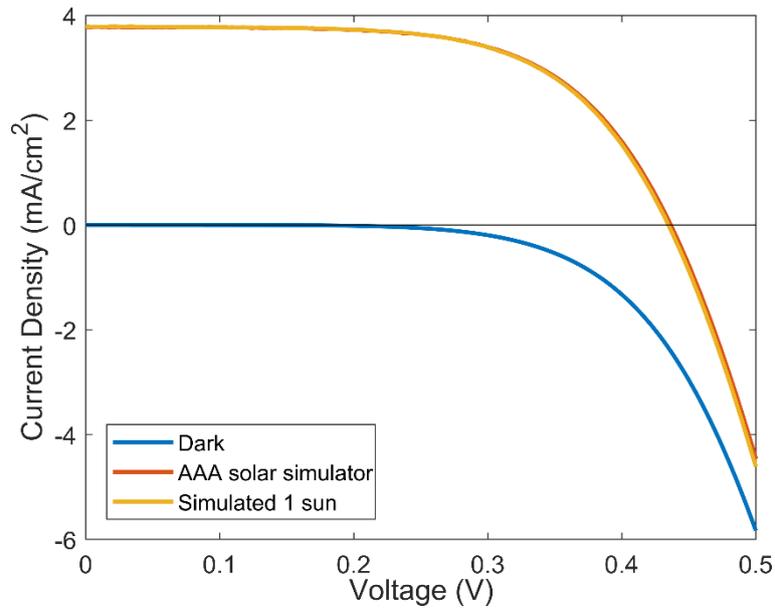

*Figure S7: Back contacts current voltage curves in the dark, and under AAA solar simulator spectrum and simulated 1 sun spectra.*

## 8. HPEV cell equivalent circuit

Equivalent circuit modeling is a widely used tool used to predict the performance of photovoltaic cells under different operating conditions.[7–10] This approach was used to estimate the HPEV electrical output with various PEC top junctions operating at different current densities. Figure S8 shows a single diode equivalent circuit as widely implemented in photovoltaic cells analysis.[11,12] As shown in the main text, the chemical output, $J_{PEC}$, is very to the back contacts operating point. Hence, assuming that the chemical output is constant, the maximum current available for extraction by the back contacts, $J_{sc}$, follows:

$$J_{sc} = q \int_0^{E_{g,PEC}} \Phi(E) \cdot EQE(E) dE - J_{PEC} \qquad (S1)$$

Where $\Phi(E)$ is the incident photons flux at energy $E$, $EQE(E)$ is the back contacts external quantum efficiency, $q$ is the elementary charge and $E_{g,PEC}$ is the bandgap of the PEC material. The current density that that can be collected at a given back contacts operating point follows the standard single diode equivalent circuit equation:

$$J_{PV} = J_{sc} - J_0 \left( \exp\left(\frac{q(V + J_{PV} \cdot R_s)}{nKT}\right) - 1 \right) \qquad (S2)$$

here $J_0$ is the back contacts diode saturation current, $J_{PV}$ and $V$ are the back contacts current density and voltage respectively, $R_s$ is the series resistance, $n$ is the diode ideality factor, $K$ is Boltzmann's constant and $T$ is the operating temperature. Hence, for a given PEC top junction band gap and current density, and a known set of solar cells parameters, the current voltage characteristics of the back contacts can be calculated from which the maximum obtainable electrical power output is extracted.

Figure 4a shows the back contacts power output at the maximum power point as a function of the PEC current density and band gap. The back contacts parameters to be used in equations ( S1 ) and ( S2 ) were extracted from EQE and current voltage data presented by Mulligan et. al.[2] The diode dark current $J_0$ is 1.2 nA/cm², the diode ideality factor $n$ is 1.475 and the series resistance, $R_s$, is 0.2Ω.

An example for the expected performance of a state of the art HPEV devices is given by combining the equivalent circuit model with published data on high efficiency PEC cells. Figure S9 shows current voltage curves of an HPEV cell as produced with the equivalent circuit and a bismuth vanadate water splitting photoanode as reported by Pihosh et. al.[5] Assuming that the HPEV front and back contacts share the same characteristics, the operating point of the coupled device can be found by intersection of the two curves. In this example, the PEC layer, which has a band gap of 2.4eV can produce hydrogen at a current of 5.6mA/cm². Thus, the short circuit current available for extraction by the back contacts is the 32mA/cm² (the entire pool of carriers available within the cell as calculated with the first term in equation ( S1 )) reduced by 5.6mA/cm² which are consumed by the chemical reaction leaving a short circuit current of 26.4mA/cm². This short circuit current is then inserted into equation ( S2 ) to produce the current voltage curve for the back contacts.

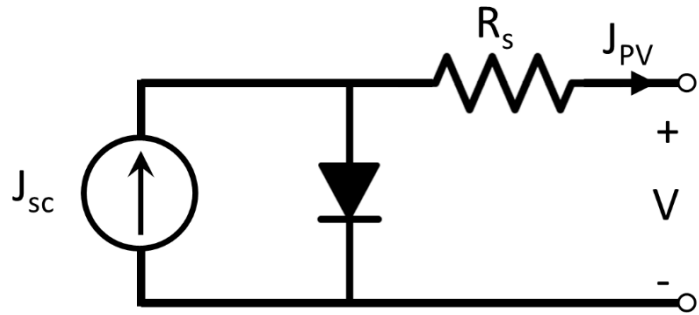

*Figure S8: The single diode equivalent circuit.*

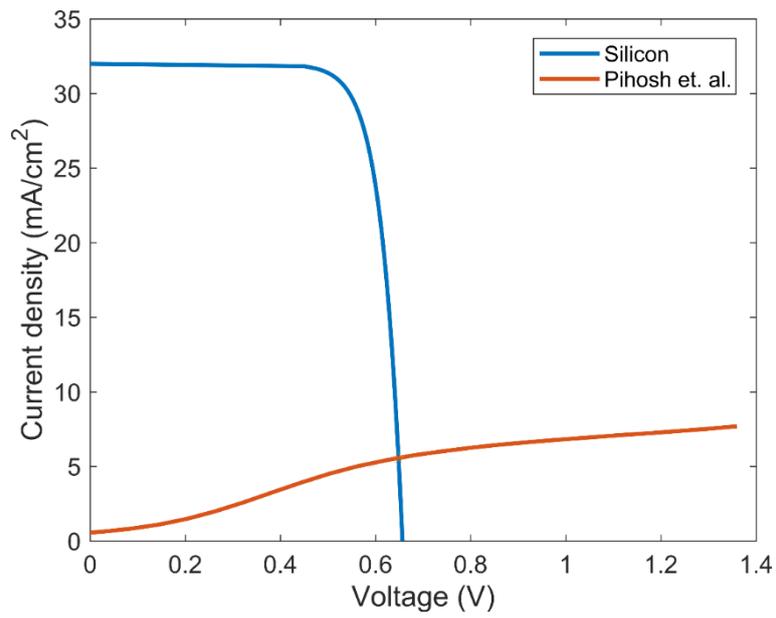

*Figure S9: The HPEV current voltage curve produced with the equivalent circuit and a bismuth vanadate PEC current voltage curve as reported by Pihosh et. al.[5]*

## 9. SEM Images of the deposited TiO$_2$ layer

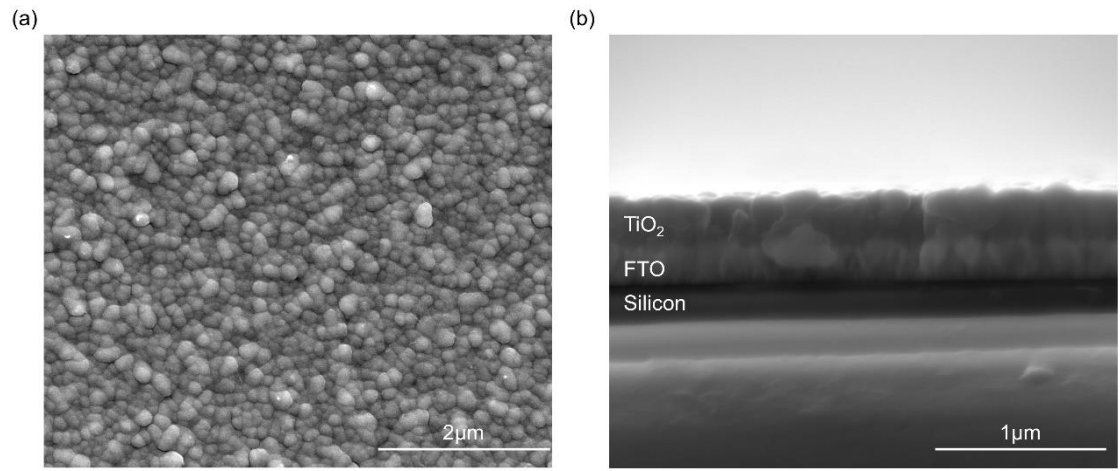

*Figure S10: plan view (a) and cross sectional SEM images of the TiO$_2$ layers photoanode deposited under the same conditions as those samples analyzed in this work.*

*10. Reflectance spectrum of a HPEV cell*

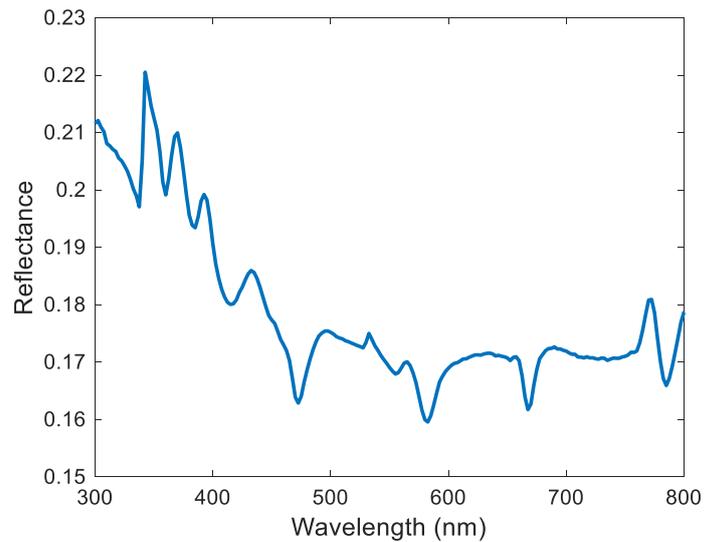

*Figure S11: the reflectance spectrum of an HPEV device fabricated with the same procedures as analyzed above.*

# References

1. Shockley, W. & Queisser, H. J. Detailed Balance Limit of Efficiency of p-n Junction Solar Cells. *J. Appl. Phys.* **32,** 510–519 (1961).

2. Mulligan, W. P. *et al.* Manufacture of solar cells with 21% efficiency. *Proc. 19th EPVSEC* 3–6 (2004).

3. Yoshikawa, K. *et al.* Silicon heterojunction solar cell with interdigitated back contacts for a photoconversion efficiency over 26%. *Nat. Energy* **2,** 17032 (2017).

4. Shaner, M. R., McDowell, M. T., Pien, A., Atwater, H. A. & Lewis, N. S. Si/TiO$_2$ Tandem-Junction Microwire Arrays for Unassisted Solar-Driven Water Splitting. *J. Electrochem. Soc.* **163,** H261–H264 (2016).

5. Pihosh, Y. *et al.* Photocatalytic generation of hydrogen by core-shell WO3/BiVO4 nanorods with ultimate water splitting efficiency. *Sci. Rep.* **5,** 11141 (2015).

6. Warren, S. C. *et al.* Identifying champion nanostructures for solar water-splitting. *Nat. Mater.* **12,** 842–9 (2013).

7. Castaner, L., Silvestre, S. & Cataluna, U. P. De. Modelling Photovoltaic Systems using PSpice.

8. Domínguez, C., Antón, I. & Sala, G. Multijunction solar cell model for translating I-V characteristics as a function of irradiance, spectrum, and cell temperature. *Prog. Photovoltaics Res. Appl.* n/a-n/a (2010). doi:10.1002/pip.965

9. Nishioka, K., Sueto, T., Uchida, M. & Ota, Y. Detailed analysis of temperature characteristics of an InGaP/InGaAs/Ge triple-junction solar cell. *J. Electron. Mater.* **39,** 704–708 (2010).

10. Walczak, K. A. *et al.* Hybrid Composite Coatings for Durable and Efficient Solar

Hydrogen Generation under Diverse Operating Conditions. *Adv. Energy Mater.* **7,** (2017).

11. Wurfel, P. *Physics of Solar Cells from Principles to New Concepts*. (Wiley, 2005).

12. Sze, S. M. & Ng, K. K. *Physics of semiconductor devices*. (Wiley-Interscience, 2007).